\documentclass[twoside,twocolumn,conference]{IEEEtran}
\usepackage{epsfig}
\usepackage{graphicx,cite,amssymb,amsmath,times}
\renewcommand{\QED}{\QEDopen}

\newcommand{\vg}[1] {{\mbox{{\boldmath ${#1}$}}}}
\newcommand{\vgs}[2] {\vg{#1}_{#2}}

\def\1N{\vgs{1}{\!N}}

\def\Pi{P_{\text{I}}}

\newtheorem{theorem}{Theorem}

\newtheorem{definition}{Definition}

\newtheorem{lemma}{Lemma}

\begin{document}
\title{Generalized Stability Condition for Generalized and Doubly-Generalized LDPC Codes}
\author{
\authorblockN{Enrico Paolini}
\authorblockA{DEIS\\
University of Bologna\\
Cesena (FC), Italy\\
epaolini@deis.unibo.it}
\and
\authorblockN{Marc Fossorier}
\authorblockA{EE Department \\
University of Hawaii at Manoa \\
Honolulu, HI 96822 \\
marc@spectra.eng.hawaii.edu}
\and
\authorblockN{Marco Chiani}
\authorblockA{DEIS\\
University of Bologna\\
Cesena (FC), Italy\\
mchiani@deis.unibo.it} }
\date{\today}
\pagestyle{empty} \thispagestyle{empty} \setcounter{page}{1}
\maketitle
%
%
\begin{abstract}
In this paper, the stability condition for low-density
parity-check (LDPC) codes on the binary erasure channel (BEC) is
extended to generalized LDPC (GLDPC) codes and doubly-generalized
LDPC (D-GLDPC) codes. It is proved that, in both cases, the
stability condition only involves the component codes with minimum
distance 2. The stability condition for GLDPC codes is always
expressed as an upper bound to the decoding threshold. This is not
possible for D-GLDPC codes, unless all the generalized variable
nodes have minimum distance at least 3. Furthermore, a condition
called derivative matching is defined in the paper. This condition
is sufficient for a GLDPC or D-GLDPC code to achieve the stability
condition with equality. If this condition is satisfied, the
threshold of D-GLDPC codes (whose generalized variable nodes have
all minimum distance at least 3) and GLDPC codes can be expressed
in closed form.
\end{abstract} {\pagestyle{plain} \pagenumbering{arabic}}
%
%
\section{Introduction}\label{section:introduction}
A traditional LDPC code \cite{gallager63:low-density} of length
$N$ and dimension $K$ is graphically represented through a
bipartite graph with $N$ variable nodes and $M \geq N-K$ check
nodes \cite{tanner81:recursive}.
%
A degree-$n$ check node of an LDPC code can be interpreted as a
length-$n$ single parity-check (SPC) code, i.e. as a $(n,n-1)$
linear block code, while a degree-$n$ variable node can be
interpreted as a length-$n$ repetition code, i.e. as a $(n,1)$
linear block code.

\begin{figure}[t]
\begin{center}
\includegraphics[width=7.5 cm, angle=0]{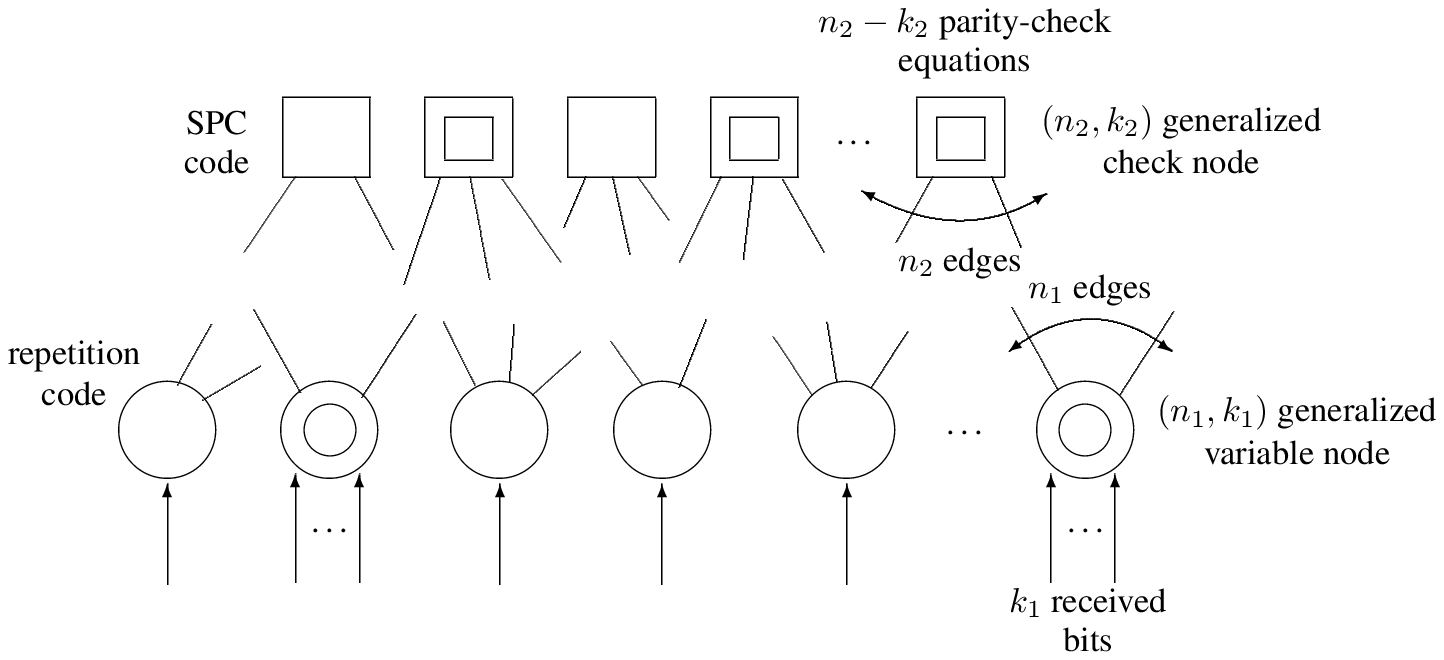}
\end{center}
\caption{Structure of a D-GLDPC code.} \label{fig:DGLDPC}
\end{figure}

Doubly-generalized LDPC (D-GLDPC) codes, recently introduced in
\cite{wang06:D-GLDPC}\cite{paolini06:doubly-allerton}, extend the
concept of LDPC codes, by allowing some variable and check nodes
to be generic linear block codes instead of repetition and SPC
codes respectively. If all the variable nodes are repetition
codes, then the code is a generalized LDPC (GLDPC) code
\cite{tanner81:recursive,boutros99:generalized,lentmaier99:generalized,fossorier05:generalized,paolini06:generalized}.
%
%
The codes used as variable and check nodes are called
\emph{component codes} of the D-GLDPC code; they will be supposed
to have minimum distance $d_{\min} \geq 2$
\cite{paolini06:doubly-allerton}. The variable and check nodes
which are not, respectively, repetition or SPC codes, are referred
to as \emph{generalized nodes}. The corresponding code structure
is represented in Fig. \ref{fig:DGLDPC}. If $N_V$ is the number of
variable nodes, then the codeword length is $N = \sum_{i=1}^{N_V}
k_i$ (with $k_i$ dimension of the $i$-th variable node). An
$(n,k)$ generalized variable node is connected to $n$ check nodes;
$k$ of the $N$ D-GLDPC encoded bits are received by the
generalized variable node, and interpreted as its $k$ information
bits. An $(n,k)$ generalized check node is connected to $n$
variable nodes. If $N_C$ is the number of check nodes, then the
number of parity check equations is $M = \sum_{i=1}^{N_C} (n_i -
k_i)$ (with $k_i$ and $n_i$, respectively, dimension and length of
the $i$-th check node). The ensembles of variable and check nodes
are called variable node decoder (VND) and check node decoder
(CND), respectively. For a description of the D-GLDPC codes
iterative decoder on the AWGN channel and the BEC, we refer to
\cite{wang06:D-GLDPC,paolini06:doubly-allerton}.

For standard LDPC ensembles, an important role is played by an
inequality known as \emph{stability condition}
\cite{richardson01:design,urbanke:modern}. For transmission on a
BEC with erasure probability $q$, the stability condition
establishes the following upper bound to the asymptotic threshold
$q^*$ for the LDPC ensemble:
\begin{align}\label{eq:stability-LDPC}
q^{*} \leq [\lambda'(0)\,\rho'(1)]^{-1}.
\end{align}
In \eqref{eq:stability-LDPC}, $\lambda'(0) = \lambda_2$ is the
fraction of edges towards the length-2 repetition variable nodes,
while $\rho'(1)$ is the derivative (computed in $x=1$) of the
function $\rho(x)=\sum_{j \geq 2}\rho_j x^{j-1}$, where $\rho_j$
is the fraction of edges connected to SPC check nodes of length
$j$. The bound \eqref{eq:stability-LDPC} was first developed from
density evolution. However, it is possible to interpret it in a
simple graphical way, by exploiting EXIT charts
\cite{ten-brink04:extrinsic}. More specifically, the stability
condition is equivalent to the following statement: 
For $q = q^*$, the derivative of the EXIT function
$I_{E,V}(I_A,q)$ for the VND, with respect to $I_A$\footnote{$I_A$
denotes the average \emph{a priori} mutual information in input to
the VND or to the CND.} and evaluated in $I_A=1$, must be smaller
than the derivative of the inverse EXIT function
$I_{E,C}^{-1}(I_A)$ for the CND, evaluated in $I_A=1$, i.e.
\eqref{eq:stability-LDPC} is equivalent to
\begin{align}\label{eq:stability-condition}
\partial I_{E,V}(I_A,q^*)/\partial I_A\,|_{I_A=1} \leq {\rm
d}I_{E,C}^{-1}/{\rm d}I_A\,|_{I_A=1}.
\end{align}

The LDPC stability condition \eqref{eq:stability-LDPC} is tight in
that there exist LDPC distributions whose threshold achieves it
with equality, assuming the closed form $q^* =
[\lambda'(0)\rho'(1)]^{-1}$. For achieving
\eqref{eq:stability-LDPC} with equality, it is sufficient that the
first occurrence of a tangency point between the EXIT function
$I_{E,V}(I_A,q)$ of the VND and the inverse EXIT function
$I_{E,C}^{-1}(I_A)$ of the CND appears in $I_A=1$, i.e.
\begin{align}\label{eq:derivative-matching}
\left\{
\begin{array}{l}
I_{E,V}(1,q^*)=I_{E,C}^{-1}(1)=1\\
\partial I_{E,V}(I_A,q^*)/\partial I_A\,|_{I_A=1} = {\rm
d}I_{E,C}^{-1}/{\rm d}I_A\,|_{I_A=1}.
\end{array}\right.
\end{align}
For LDPC codes, the first equality is always true. As proved in
\cite{paolini06:doubly-allerton}, it is always satisfied also for
GLDPC and D-GLDPC codes, if all the variable and check component
codes have $d_{\min} \geq 2$, which is assumed true in this paper.
Then, only the second equality will be considered in the sequel,
and referred to as \emph{derivative matching condition}.

In this paper, the stability condition
\eqref{eq:stability-condition}, and the derivative matching
condition \eqref{eq:derivative-matching} are extended to GLDPC and
D-GLDPC codes. Two main results are obtained. The first is that
only the component codes with $d_{\min}=2$, including length-2
repetition codes and SPC codes, appear in the stability condition.
The second is that, for GLDPC codes satisfying the derivative
matching condition, it is always possible to develop a closed-form
expression of the threshold; the same expression holds also for
D-GLDPC codes satisfying the derivative matching condition, if all
the generalized variable nodes have $d_{\min} \geq 3$.
\section{Definitions and Basic Notation}\label{section:definitions}
The transmission channel is a BEC with erasure probability $q$.
Assuming a bipartite graph with random connections, the
\emph{extrinsic channel}, over which the messages are exchanged
between the variable and check nodes, during the iterative
decoding process, is modelled as a second BEC with erasure
probability $p$ \cite{ten-brink04:extrinsic} (depending on the
decoding iteration). It is readily proved that $I_A = 1-p\,$:
Since the EXIT functions will be expressed as functions of $p$
(and $q$ for the VND), the derivatives of the VND EXIT function
and of the CND inverse EXIT function will be evaluated at $p=0$
($I_A=1$).

Under the hypothesis of random bipartite graph, the VND and CND
EXIT functions can be expressed, respectively, as
\begin{align}\label{eq:exit-VND}
I_{E,V}(p,q) =
\sum_{i=1}^{\mathcal{I}_V}\lambda_i\,I_{E,V}^{(i)}(p,q)
\end{align}
\begin{align}\label{eq:exit-CND}
I_{E,C}(p) = \sum_{i=1}^{\mathcal{I}_C}\rho_i\,I_{E,C}^{(i)}(p),
\end{align}
where $\mathcal{I}_V$ and $\mathcal{I}_C$ are the number of
variable and check node types, $I_{E,V}^{(i)}(p,q)$ and
$I_{E,C}^{(i)}(p)$ are the EXIT function for the $i$-th variable
node type and $i$-th check node type, $\lambda_i$ and $\rho_i$ are
the fractions of edges towards the variable nodes of type $i$ and
the check nodes of type $i$.

For the scope of this work it is useful to isolate, in
\eqref{eq:exit-VND}, the contribution of the repetition codes and,
in \eqref{eq:exit-CND}, the contribution of the SPC codes:
\begin{align}\label{eq:exit-VND-divide}
I_{E,V}(p,q) & = \sum_{j \geq 2}^{\rm (rep)}\lambda_j^{\rm (r)
}\cdot(1-q\,p^{j-1}) + \sum_{i}^{\rm
(gen)}\lambda_i\,I_{E,V}^{(i)}(p,q) \notag \\
\, & = \sum_{j \geq 2}^{\rm (rep)}\lambda_j^{\rm (r)} -
q\,\lambda_{\rm r}(p) + \sum_{i}^{\rm
(gen)}\lambda_i\,I_{E,V}^{(i)}(p,q)
\end{align}
\begin{align}\label{eq:exit-CND-divide}
I_{E,C}(p) & = \sum_{j \geq 2}^{\rm (SPC)}\rho_j^{\rm (SPC)
}\cdot(1-p)^{j-1} +
\sum_{i}^{{\rm(gen)}}\rho_i\,I_{E,C}^{(i)}(p) \notag \\
\, & = \rho_{SPC}(1-p) +
\sum_{i}^{{\rm(gen)}}\rho_i\,I_{E,C}^{(i)}(p).
\end{align}
In \eqref{eq:exit-VND-divide}, $j$ is the length of the generic
repetition variable node, $\lambda_j^{\rm (r)}$ is the fraction of
edges towards the repetition nodes of length $j$, $\lambda_{\rm
r}(x)$ is defined as $\sum_{j \geq 2} \lambda_j^{\rm
(r)}\,x^{j-1}$. It uses the well known EXIT function expression on
the BEC, $I_E(p,q) = 1 - q \, p^{j-1}$, for a $(j,1)$ repetition
variable node. The summation in $i$ is over all the generalized
variable node types. Analogously, in \eqref{eq:exit-CND-divide},
$j$ is the length of the generic SPC node, $\rho_j^{\rm (SPC)}$ is
the fraction of edges towards the SPC nodes of length $j$,
$\rho_{\rm SPC}(x)$ is defined as $\sum_{j \geq 2} \rho_j^{\rm
(SPC)}\,x^{j-1}$, and the expression $I_E(p) = (1-p)^{j-1}$, valid
for a $(j\,,j-1)$ SPC check node, is used.

The EXIT function for an $(n,k)$ generalized variable node of a
D-GLDPC code on the BEC can be expressed as
\begin{align}\label{eq:modified-IE-variable}
I_E(p,q) = 1 & - \frac{1}{n} \sum_{t=0}^{n-1}\,\sum_{z=0}^{k}
a_{t,z}\,p^t\,(1-p)^{n-t-1}\,q^z\,(1-q)^{k-z},
\end{align}
expression which can be developed from \cite[eq.
36]{ten-brink04:extrinsic}, where $a_{t,z} =
[(n-t)\widetilde{e}_{n-t,k-z}-(t+1)\widetilde{e}_{n-t-1,k-z}]$.
The parameter $\tilde{e}_{g,h}$ (with $g=0,\dots,n$ and
$h=0,\dots,k$) is the $(g,h)$-th un-normalized split information
function, defined as explained next. Considering a representation
of the generator matrix $\mathbf{G}$ for the $(n,k)$ variable
node, and appending to it the $(k \times k)$ identity matrix
$\mathbf{I}_k$, $\tilde{e}_{g,h}$ is equal to the summation of the
ranks over all the possible submatrices obtained selecting $g$
columns in $\mathbf{G}$ and $h$ columns in $\mathbf{I}_k$. We
remark that the split information functions for a generalized
variable node, and then its EXIT function, heavily depend on the
code representation, i.e. on the chosen generator matrix
\cite{paolini06:doubly-allerton}. Then, the performance of the
overall D-GLDPC code depends on the code representation used at
the generalized variable nodes.

The EXIT function for a generalized $(n,k)$ check node of a GLDPC
or D-GLDPC code on the BEC can be obtained by letting $q
\rightarrow 1$ in \eqref{eq:modified-IE-variable} (no
communication channel is present). The obtained expression,
equivalent to \cite[eq. 40]{ten-brink04:extrinsic}, is
\begin{align}\label{eq:modified-IE-check}
I_E(p) = 1 - \frac{1}{n} \sum_{t=0}^{n-1} a_t p^{t}(1-p)^{n-t-1},
\end{align}
where, $a_t = (n-t)\,\tilde{e}_{n-t}-(t+1)\tilde{e}_{n-t-1}$ and
where for $g=0,\dots,n$, $\tilde{e}_g$ is the $g$-th un-normalized
information function of the $(n,k)$ code, a concept first
introduced in \cite{helleseth97:information}. It is defined as the
summation of the ranks over all the possible submatrices obtained
selecting $g$ columns from the generator matrix $\mathbf{G}$. As
opposed to the split information functions, the information
functions are independent of the code representation. Thus,
different code representations lead to the same EXIT function for
the generalized check node. The performance of the overall D-GLDPC
code is then independent of the specific representation of the
generalized check nodes.

Equations \eqref{eq:modified-IE-variable} and
\eqref{eq:modified-IE-check} assume that MAP erasure correction is
performed at the variable and check node.
\section{Independent Sets and Minimum Distance}\label{section:dmin}
The development of a generalized stability condition for GLDPC and
D-GLDPC codes is mostly based on a theorem proposed in this
section. This theorem establishes a sufficient condition for a $(k
\times (n-t))$ binary matrix, obtained by selecting $n-t$ columns
in the generator matrix $\mathbf{G}$ (any representation) of a
$(n,k)$ linear block code, to have rank equal to $k$. The results
presented in this section can be also developed from
\cite{helleseth97:information}. Here, they are independently
proved and formulated for the scope of the paper.
\medskip
\begin{definition}[independent set]
Given a $(k \times n)$ binary matrix of rank $r$, a set of $t$
columns is said an \emph{independent set} when removing these $t$
columns from the matrix leads to a $(k \times (n-t))$ matrix with
rank $r - \Delta r< r$, for some $0 < \Delta r \leq t$. The number
$t$ is the size of the independent set.
\end{definition}
\medskip

If $j$ columns are an independent set for a certain representation
of the generator matrix, they are an independent set for any other
representation. Moreover, removing them from any representation of
the generator matrix leads to the same rank reduction $\Delta r$.
This is because all the possible representations of the generator
matrix can be obtained by row summations starting from any other
representation.
\medskip
\begin{lemma}\label{lemma:dmin-bound}
If any representation of the generator matrix of a $(n,k)$ linear
block code has an independent set of size $j$, then the code
minimum distance satisfies $d_{\min} \leq j$.
\end{lemma}
\begin{figure}[t]
\begin{center}
\includegraphics[width=6 cm, angle=0]{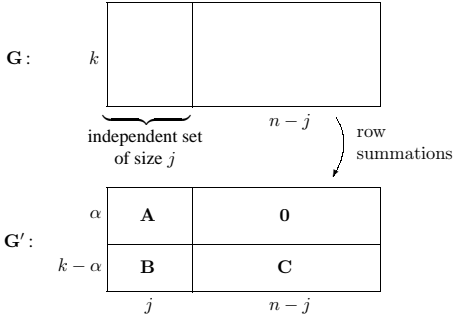}
\end{center}
\caption{If ${\rm rank}(\mathbf{G})=r$ and the first $j$ columns
are an independent set, then $\mathbf{G}$ can be transformed as
shown by row summations only, with $\mathbf{A}$, $\mathbf{B}$ and
$\mathbf{C}$ non-null matrices.} \label{fig:row_sum}
\end{figure}
\begin{proof}
Let us suppose that $j$ columns of a generator matrix $\mathbf{G}$
are an independent set of size $j$. Then, it must be possible to
perform row summations on $\mathbf{G}$ in order to obtain a new
generator matrix representation $\mathbf{G}'$, in which a certain
number $\alpha$ of rows have all their 1's in correspondence of
only the columns of the independent set (see for example Fig.
\ref{fig:row_sum}, where the first $j$ columns are supposed an
independent set, and where $\mathbf{A}$, $\mathbf{B}$ and
$\mathbf{C}$ are non-null matrices). Any of these $\alpha$ rows is
a valid codeword. Then, $d_{\min} \leq j$.
\end{proof}

\medskip
\begin{theorem}\label{theorem:min-indep-set-size}
Let $\mathbf{G}$ be any representation of the generator matrix of
an $(n,k)$ linear block code. Then, the following statements are
equivalent:
\begin{itemize}
\item[a)] the code has minimum distance $t$;
\item[b)] the minimum size of the independent sets of $\mathbf{G}$
is $t$.
\end{itemize}
\end{theorem}
\begin{proof}
$[{\rm a}\,\Rightarrow\,{\rm b}]$ If $d_{\min} = t$, then it is
possible to construct a representation of $\mathbf{G}$ where there
is at least one row with exactly $t$ 1's. The columns of
$\mathbf{G}$ corresponding to these $t$ 1's are an independent set
(of size $t$), because removing them from $\mathbf{G}$ leads to a
reduction of the rank. This independent set must be of minimum
size. In fact, if it existed an independent set of size $j < t$,
then from Lemma \ref{lemma:dmin-bound} it would follow $d_{\min} <
t$, thus violating the hypothesis $d_{\min} = t$.\\
$[{\rm b}\,\Rightarrow\,{\rm a}]$ Let us suppose that the minimum
size of the independent sets of $\mathbf{G}$ is $t$, and let us
consider an independent set of size $t$. From Lemma
\ref{lemma:dmin-bound} it follows that $d_{min} \leq t$. The proof
is completed by showing that it is not possible to have $d_{\min}
< t$. In fact, if $d_{\min} = j < t$ then, by reasoning in the
same way as for the $[{\rm a}\,\Rightarrow\,{\rm b}]$ proof, it
would follow that the minimum size of the independent sets of
$\mathbf{G}$ is $j < t$, which violates the hypothesis.
\end{proof}
\medskip
The present section is concluded by the following lemma (not
necessary to prove Theorem \ref{theorem:min-indep-set-size}).
\medskip
\begin{lemma}
Let $k < n$, and $t$ be the minimum size of the independent sets
of a $(k \times n)$ binary matrix with rank $r$. Then, removing
any independent set of size $t$, leads to a $(k \times (n-t))$
matrix with rank $r-1$.
\end{lemma}
\begin{proof}
Since $t$ is the minimum size of the independent sets of the
matrix, then removing any set of $j < t$ columns does not affect
the rank. For an independent set of size $t$, one can remove any
subset of $t-1$ columns without reducing the rank; when removing
the $t$-th column, the rank can only decrease by 1.
\end{proof}
%
\section{Stability Condition and Derivative Matching for GLDPC Codes}\label{section:gldpc}
In GLDPC codes, all the variable nodes are repetition codes.
Recalling \eqref{eq:exit-VND-divide}, and observing that in this
case $\sum_{j \geq 2}^{\rm (rep)}\lambda_j^{\rm (r)}=1$, the EXIT
function on the BEC for the VND is given by $I_{E,V}(p,q) = 1 - q
\lambda^{(\rm r)}(x)$. Hence:
\begin{align}\label{eq:derivative-variable-gldpc}
\partial I_{E,V}(p,q) / \partial p \,|_{p=0} = -q \,
\lambda^{(\rm r)}_2.
\end{align}

Recalling \eqref{eq:exit-CND-divide}, the derivative of
$I_{E,C}(p)$ at $p=0$ is
\begin{align}\label{eq:exit-CND-derivative}
\frac{{\rm d} I_{E,C}(p)}{{\rm d} p} \,\Big|_{p=0} = -\rho'_{\rm
SPC}(1) + \sum_{i}^{{\rm(gen)}}\rho_i\,\frac{{\rm
d}I_{E,C}^{(i)}(p)}{{\rm d} p}\Big|_{p=0}.
\end{align}

In order to develop the previous expression, it is necessary to
express the derivative of the EXIT function for the generalized
check nodes. This task can be performed by exploiting the theorem
proved in Section \ref{section:dmin}, as explained next.

Consider an $(n,k)$ generalized check node, with EXIT function
$I_E(p)$ in the form \eqref{eq:modified-IE-check}. It is readily
shown that the derivative of $I_E(p)$, computed at $p=0$, is
$ \frac{{\rm d} I_E(p)}{{\rm d} p} \,|_{p=0} = \frac{(n-1)a_0 -
a_1}{n} $.
According to Theorem \ref{theorem:min-indep-set-size}, $a_0 = 0$
if and only if the generalized check node has minimum distance
$d_{\min} \geq 2$. In fact, the generator matrix of the check node
is full rank (rank $= k$) by definition, so $\tilde{e}_n = k$.
Furthermore, from Theorem \ref{theorem:min-indep-set-size},
removing any single column from the generator matrix does not
reduce the rank if and only if $d_{\min} \geq 2$, thus leading to
$\tilde{e}_{n-k} = n\,k$. Then, $a_0 = n\,\tilde{e}_n -
\tilde{e}_{n-1} = n\,k - n\,k = 0$. As recalled in Section
\ref{section:introduction}, the hypothesis $d_{\min} \geq 2$ is
always assumed in this paper. Then, it will be always assumed $a_0
= 0$.

If $d_{\min} \geq 2$ for the check node, then
$ \frac{{\rm d} I_E(p,q)}{{\rm d} p} \,|_{p=0} = - \frac{a_1}{n},
$
where $a_1 = (n-1)\tilde{e}_{n-1} - 2\,\tilde{e}_{n-2} =
k\,n\,(n-1)- 2\,\tilde{e}_{n-2}$. By applying again Theorem
\ref{theorem:min-indep-set-size}, we obtain
\begin{align}\label{eq:a1-expression}
a_1 \left\{ \begin{array}{l}
= 0 \quad {\rm if } \,\,d_{\min} \geq 3\\
\neq 0 \quad {\rm if } \,\,d_{\min} = 2\,. \end{array} \right.
\end{align}
In fact, if $d_{\min} \geq 3$, removing any pair of columns from
the generator matrix does not affect the rank. In this case
$2\,\tilde{e}_{n-2} = 2 \, k\,{n \choose 2} = k\,n\,(n-1)$, hence
$a_1 = 0$.

According to these results, the only generalized check nodes that
give some contribution to the summation in the second member of
\eqref{eq:exit-CND-derivative} are those characterized by
$d_{\min} = 2$. By recalling that all the SPC codes have minimum
distance 2, we conclude that $\frac{{\rm d} I_{E,C}(p)}{{\rm d} p}
\,|_{p=0}$ only depends on the check nodes with $d_{\min} = 2$.
The derivative at $p=0$ of the EXIT function for the CND can be
then expressed as
\begin{align}\label{eq:exit-CND-deriv-d2}
&\frac{{\rm d} I_{E,C}(p)}{{\rm d} p} \,\Big|_{p=0} \notag \\
\, & = -\rho'_{\rm SPC}(1) -
\sum_{i}^{d_{\min}=2}\rho_i\,\frac{k_i n_i
(n_i-1)-2\,\tilde{e}_{n_i-2}}{n_i} \notag \\
\, & = -\rho'_{\rm SPC}(1) -
\sum_{i}^{d_{\min}=2}\frac{2\rho_i}{n_i}\,\Delta_{n-2}^{(i)},
\end{align}
where the summation is over the generalized check node types with
minimum distance 2. In the last equality, $\Delta_{n-2}^{(i)} > 0$
is defined as follows. Let $\mathcal{S}_{n_i-2}$ be the generic
$(k_i \times (n_i-2) )$ matrix obtained by selecting $n_i - 2 $
columns in the generator matrix. Then, $\Delta_{n-2}^{(i)} =
\sum_{\mathcal{S}_{n_i-2}} (k_i - {\rm
rank}(\mathcal{S}_{n_i-2}))$, where the summation is over all the
possible matrices $\mathcal{S}_{n_i-2}$. The parameter
$\Delta_{n-2}^{(i)}$ does not depend on the chosen representation
for the $i$-th generalized check node.

The derivative at $p=0$ for the inverse EXIT function
$I_{E,C}^{-1}(p)$ of the CND is simply given by the inverse of
\eqref{eq:exit-CND-deriv-d2}. Then, the stability condition
$\partial I_{E,V}(I_A,q^*)/\partial I_A\,|_{I_A=1} \leq {\rm
d}I_{E,C}^{-1}/{\rm d}I_A\,|_{I_A=1}$ for GLDPC codes leads to %
\footnote{If the derivatives are computed with respect to
$p=1-I_A$, the stability condition is $\partial
I_{E,V}(p,q^*)/\partial p\,|_{p=0} \geq {\rm d}I_{E,C}^{-1}/{\rm
d}p\,|_{p=0}$.}
\begin{align}\label{eq:stability-GLDPC}
q^* \leq \Big[ \lambda^{(\rm r)}_2 \Big( \rho'_{\rm SPC}(1) +
\sum_{i}^{d_{\min}=2}\frac{2\rho_i}{n_i}\,\Delta_{n-2}^{(i)} \Big)
\Big]^{-1},
\end{align}
an upper bound on the threshold $q^*$ which represents a necessary
condition for successful GLDPC (asymptotic) decoding\footnote{Some
results about the stability condition on the BEC for GLDPC codes
can be also found in \cite[Appendix 7.A]{measson:2006phd}.}.

For GLDPC codes satisfying the derivative matching condition
\eqref{eq:derivative-matching} (the first occurrence of a tangency
point between $I_{E,V}(p,q)$ and $I_{E,C}^{-1}(p)$ appears at
$p=0$), the threshold assumes the following simple closed form:
\begin{align}\label{eq:q*-GLDPC}
q^* = \Big[ \lambda^{(\rm r)}_2 \Big( \rho'_{\rm SPC}(1) +
\sum_{i}^{d_{\min}=2}\frac{2\rho_i}{n_i}\,\Delta_{n-2}^{(i)} \Big)
\Big]^{-1}.
\end{align}

If only generalized check nodes with $d_{\min} \geq 3$ are used,
the stability condition \eqref{eq:stability-GLDPC} and the
threshold expression \eqref{eq:q*-GLDPC} become, respectively,
\begin{align}\label{eq:der-match-GLDPC-dmin>3}
q^* \leq \Big[ \lambda^{(\rm r)}_2 \, \rho'_{\rm SPC}(1)
\Big]^{-1}
\end{align}
\begin{align}\label{eq:stability-GLDPC-dmin>3}
q^* = \Big[ \lambda^{(\rm r)}_2 \, \rho'_{\rm SPC}(1) \Big]^{-1}.
\end{align}

\section{Stability Condition and Derivative Matching for D-GLDPC Codes}\label{section:d-gldpc}
The derivative at $p=0$ of the EXIT function for the CND of
D-GLDPC codes is the same as for GLDPC codes, and is expressed by
\eqref{eq:exit-CND-deriv-d2}. The derivative with respect to $p$,
at $p=0$, of the EXIT function for the VND is developed next.

It follows from \eqref{eq:exit-VND-divide}:
\begin{align}\label{eq:derivative-EXIT-variable-div}
\frac{\partial I_{E,V}(p,q)}{\partial p} \Big|_{p=0} =  -
q\,\lambda^{\rm (r)}_2 + \sum_{i}^{\rm (gen)}\lambda_i\,
\frac{\partial I_{E,V}^{(i)}(p,q)}{\partial p} \Big|_{p=0}.
\end{align}
In order to develop the summation on the generalized variable node
types in the second term, the derivative of the EXIT function for
each generalized variable node type, with respect to $p$ and at
$p=0$, can be computed based on \eqref{eq:modified-IE-variable}.
By defining $f(p) =
\sum_{t=0}^{n-1}\,a_{t,z}\,p^t\,(1-p)^{n-1-t}$, it results
\begin{align}
\frac{\partial I_E(p,q)}{\partial p} & \Big|_{p=0} = -\frac{1}{n}
\, \sum_{z=0}^{k}\,q^z\,(1-q)^{k-z} \frac{{\rm d}f(p)}{{\rm d}
p}\Big|_{p=0} \notag \\
\, & = \sum_{z=0}^{k}\,q^z\,(1-q)^{k-z}\,\frac{(n-1)a_{0,z} -
a_{1,z}}{n},
\end{align}
where the fact ${\rm d}f(p)/{\rm d}p\,|_{p=0} = -(n-1)a_{0,z} +
a_{1,z}$ has been exploited. The previous relationship can be
further developed by exploiting Theorem
\ref{theorem:min-indep-set-size}. Since, by hypothesis, any
variable node has minimum distance at least 2, removing any single
column from the generator matrix $\mathbf{G}$ of the code
associated to the variable node does not affect the rank of
$\mathbf{G}$. It follows
$ a_{0,z} = n\, \widetilde{e}_{n,k-z}-\widetilde{e}_{n-1,k-z} =
k\,n\,{k \choose k-z} - k\,n\,{k \choose k-z} = 0 $,
thus leading to
$ \frac{\partial I_E(p,q)}{\partial p}|_{p=0} = -\,
\sum_{z=0}^{k}\,q^z\,(1-q)^{k-z} \, \frac{a_{1,z}}{n} $.

Theorem \ref{theorem:min-indep-set-size} can be invoked again in
order to show that
\begin{align}\label{eq:a1z-expression}
a_{1,z} \left\{ \begin{array}{l}
= 0 \quad \forall \, z \quad {\rm if } \,\,d_{\min} \geq 3\\
\neq 0 \quad \forall \, z \quad {\rm if } \,\,d_{\min} = 2,
\end{array} \right.
\end{align}
where $d_{\min}$ is the minimum distance of the $(n,k)$ code
associated to the variable node under analysis. In fact, under the
hypothesis $d_{\min} \geq 3$, no independent sets of size 1 and 2
are present in (any representation of) the generator matrix
$\mathbf{G}$. Then
$ a_{1,z} = (n-1)\,\tilde{e}_{n-1,k-z} - 2\,\tilde{e}_{n-2,k-z} =
k\,n\,(n-1)\,{k \choose k-z} - 2\,k\,{n \choose n-2}\,{k \choose
k-z} = 0 $.

It follows that the contribution of the generalized variable nodes
to $\frac{\partial I_{E,V}(p,q)}{\partial p}|_{p=0}$ is 0 if they
all have minimum distance greater than or equal to 3. The only
non-null contribution comes from the generalized variable nodes
with minimum distance $d_{\min}=2$, which is coherent with the
fact that, among the repetition codes, only those with
$d_{\min}=2$ (i.e. the length-2 repetition codes) give a non-null
contribution.

Then, \eqref{eq:derivative-EXIT-variable-div} can be developed as
\begin{align}\label{eq:derivative-EXIT-variable-d2}
\, & \frac{\partial I_{E,V}(p,q)}{\partial p} \Big|_{p=0} =  -
q\,\lambda^{\rm (r)}_2
-\sum_{i}^{d_{\min}=2}\,\sum_{z=0}^{k_i}\,q^z\,(1-q)^{k_i-z} \notag \\
\, & \phantom{------} \cdot \lambda_i
\frac{k_i\,n_i(n_i-1){k_i \choose k_i-z}-2\,\tilde{e}_{n_i-2,k_i-z}}{n_i} \notag \\
\, & = - q\,\lambda^{\rm (r)}_2
-\sum_{i}^{d_{\min}=2}\,\sum_{z=0}^{k_i}\,q^z\,(1-q)^{k_i-z}
\frac{2\lambda_i}{n_i}\Delta_{n-2,k-z}^{(i)}.
\end{align}
In the previous expression, $\Delta_{n-2,k-z}^{(i)}$ is defined in
a similar way as $\Delta_{n-2}^{(i)}$ for GLDPC codes. Let
$\mathcal{S}_{n_i-2,k_i-z}$ be the generic $(k_i \times
(n_i-2+k_i-z) )$ matrix obtained by selecting $n_i - 2 $ columns
in the generator matrix, and $k_i-z$ columns in the $(k_i \times
k_i)$ identity matrix. Then, $\Delta_{n-2,k-z}^{(i)} =
\sum_{\mathcal{S}_{n_i-2,k_i-z}} (k_i - {\rm
rank}(\mathcal{S}_{n_i-2,k_i-z}))$, where the summation is over
all the possible matrices $\mathcal{S}_{n_i-2,k_i-z}$. Differently
from $\Delta_{n-2}^{(i)}$, the parameter $\Delta_{n-2,k-z}^{(i)}$
depends on the code representation.

By combining \eqref{eq:exit-CND-deriv-d2} and
\eqref{eq:derivative-EXIT-variable-d2} the stability condition
$\partial I_{E,V}(p,q^*) / \partial p|_{p=0} \geq {\rm d}
I_{E,C}^{-1}(p) / {\rm d}p \, |_{p=0}$ assumes the form:
\begin{align}\label{eq:stability-dgldpc}
\, & q^*\,\lambda^{\rm (r)}_2
+\sum_{i}^{d_{\min}=2}\,\sum_{z=0}^{k_i}\,(q^*)^z\,(1-q^*)^{k_i-z}
\frac{2\lambda_i\,\Delta_{n-2,k-z}^{(i)}}{n_i} \notag \\
\, & \phantom{----} \leq \Big[ \rho'_{\rm SPC}(1) +
\sum_{i}^{d_{\min}=2}\frac{2\rho_i}{n_i}\,\Delta_{n-2}^{(i)}
\Big]^{-1}.
\end{align}

This inequality is a necessary condition for (asymptotic)
successful D-GLDPC decoding on the BEC. Differently from the GLDPC
case, it is not possible to express this inequality as an upper
bound on $q^*$, because of the impossibility to factor $q^*$ from
the summation in $z$. For D-GLDPC codes satisfying the derivative
matching condition, \eqref{eq:stability-dgldpc} holds with
equality. For the same reason, it does not lead to an explicit
closed form expression of the threshold. However, if $d_{\min}
\geq 3$ for all the generalized variable nodes, then any term of
the summation over the generalized variable nodes is null. In this
case, the same upper bound to $q^*$ as in \eqref{eq:q*-GLDPC}
holds:
$$
q^* \leq \Big[ \lambda^{(\rm r)}_2 \Big( \rho'_{\rm SPC}(1) +
\sum_{i}^{d_{\min}=2}\frac{2\rho_i}{n_i}\,\Delta_{n-2}^{(i)} \Big)
\Big]^{-1},
$$
and this inequality is achieved with equality by D-GLDPC codes
satisfying the derivative matching condition. Moreover, if
$d_{\min} \geq 3$ also for all the generalized check nodes, then
the upper bound on $q^*$ assumes the simple form as in
\eqref{eq:der-match-GLDPC-dmin>3}:
\begin{align*}
q^* \leq \Big[ \lambda^{(\rm r)}_2 \, \rho'_{\rm SPC}(1)
\Big]^{-1}.
\end{align*}
This inequality is achieved with equality by D-GLDPC codes
satisfying the derivative matching condition.
\section{Conclusion}\label{section:conclusion}
In this paper, a stability condition on the BEC has been derived
for GLPDC and D-GLDPC codes, generalizing the inequality $q^* \leq
[\lambda'(0)\rho'(1)]^{-1}$, valid for standard LDPC codes. A
derivative matching condition has been also defined, sufficient
to achieve the stability condition with equality.

As for LDPC codes, only the variable and check nodes with minimum
distance 2 are involved in the stability condition. 
The stability condition for GLDPC codes can be always explicitly
expressed as an upper bound on the decoding threshold. D-GLDPC
codes do not share this property in general; however, if all the
generalized variable nodes have minimum distance at least 3, the
stability condition becomes the same as for GLDPC codes. As a
consequence, for GLDPC codes and for D-GLDPC codes with variable
component codes of minimum distance at least 3 the decoding
threshold assumes a simple closed form.

\section*{Acknowledgment}
This work has been supported in part by NSF under Grant
CCF-0515154 and in part by ESA/ESOC.
%
%
%
%

\end{document}